\documentstyle{article}
\title{(1,0) superparticle in a curved background
and the full set of (1,0) supergravity constraints}
\author{A.A. Deriglazov\thanks{E-mail: theordpt@fftgu.tomsk.su}\\
Department of Mathematical Physics,\\
Politechnical University, Tomsk, 634055, Russia\\
and\\
A.V. Galajinsky\\
Department of Theoretical Physics,\\
Tomsk State University, Tomsk, 634050, Russia}
\date{}

\begin{document}
\maketitle

\abstract{The classical (1,0) superparticle in a curved superspace is
considered. The minimal set of constraints to be imposed on the
background for correct inclusion of interaction is found. The most
general form of Siegel-type local fermionic symmetry is presented.
Local algebra of the theory is shown to be closed off-shell and
nontrivially deformed as compared to the flat one. The requirements
leading to the full set of (1,0) supergravity constraints are
presented.}

\section{Introduction}

In recent years, propagation of the superparticle [1] and of the superstring
[2] in the presence of external background superfields was intensively
investigated [3--7]. It was shown in the original paper [3] that the
requirement of the Siegel's local fermionic symmetry [8] in the $d=10$
superparticle action (correct inclusion of interaction has to preserve
local symmetries of a theory) led to some constraints on a background.
Namely, the full set of SYM constraints arised in the case of SYM
background superfield and some part of the full set of supergravity
constraints appeared in the case of a curved background. Different
attempts to get the full set of supergravity constraints proceeding from the
consideration of the superparticle in a curved background were proposed
in Refs. [4, 9].

The straightforward way to get the superparticle model in a curved
background includes three steps [3]. First of all, the flat action and
gauge transformations are to be written in terms of the flat
supervielbein. Second, the coordinates and supervielbein are set to be
the coordinates and supervielbein of a curved superspace. Third, since
the resulting action doesn't possess any local symmetry except the
reparametrization invariance, it is necessary to demand the
$k$-transfor\-mations (arising at the first and second steps) to be a
symmetry of the superparticle action. One can note, however, that this
procedure doesn't guarantee that the arising in such a way gauge
transformations are of the most general form. In the present paper we
consider the (1,0) superparticle in (1,0) curved superspace.
Proceeding from the minimal set of constraints which must be
imposed on background geometry for correct inclusion of
interaction, we find the most general form of Siegel's local fermionic
transformations for the theory. An algebraic structure of the
arising transformations is investigated and requirements leading
to the full set of (1,0) supergravity constraints are presented.

We begin with brief consideration of the model in flat superspace in
Sec. 2. A full gauge algebra of the theory is shown to be closed
off-shell, as opposed to the case of another dimensions. In Sec. 3
we examine the model in curved superspace within the Hamiltonian
formalism and find the minimal set of constraints on background
necessary for correct inclusion of interaction. We do these using
the Dirac's constraint formalism and requiring a correct number
of first- and second-class constraints for the model. Our analysis here
is closely related to that of Ref. [4]. Next, in the Lagrangian
formalism we reconstruct the most general form of Siegel's
transformations which are consistent with these restrictions on the
background. It is interesting to note that the arising transformations
contain non-trivial contributions including the torsion superfield and
don't coincide with the direct generalization of the flat Siegel's
transformations. In Sec. 4 full gauge algebra is evaluated and shown to
be closed off-shell and nontrivially deformed as compared to the flat
one. Further, we construct the formulation leading to the full set of
(1,0) supergravity constraints. The full set follows from
requirement that the direct generalization of the flat gauge
transformations is realized in the model. One can note the similarity
with the (1,0) chiral superparticle [10]. In the conclusion the
possibilities of extending the results to the case of another dimensions
are discussed.

\section{Off-shell closure of local symmetries}

We use the real representation for $\Gamma^M$-matrices in $d=2$ and
light-cone coordinates for bosonic variables. (1,0) superspace is
parametrized by the coordinates $z^M=(x^+,x^-,\theta^\alpha )$, where
$\theta^\alpha \equiv \left(\begin{array}{l}\theta^{(+)}\\
0 \end{array} \right)$ is a Majorana--Weyl spinor. In these notations
the action for the (1,0) superparticle in flat superspace has the
following form:
$$
S = \int {\rm d}\tau \,e^{-1}\Pi^+{\dot x}^-
\eqno{(1)}$$
where we denoted $\Pi^+={\dot x}^+-{\rm i}\dot\theta^{(+)}
\theta^{(+)}$ and $e$ is an auxiliary einbein field. The theory possesses
global (1,0)-supersymmetry
$$
\delta\theta^{(+)}=\epsilon^{(+)}, \qquad \delta x^+ = {\rm i}
\theta^{(+)}\epsilon^{(+)}.
\eqno{(2)}$$
The local symmetries of the model include the standard
reparametrization invariance with a bosonic parameter $\alpha (\tau )$.
$$
\begin{array}{l} \delta_\alpha x^+ =\alpha{\dot x}^+, \qquad
\delta_\alpha x^- =\alpha{\dot x}^-,\\
\delta_\alpha \theta^+ =\alpha{\dot\theta}^+, \qquad \delta_\alpha e
=(\alpha e)^{\cdot}.\end{array}
\eqno{(3)}$$
and the Siegel's transformations with a fermionic parameter $k^{(-)}
(\tau )$
$$
\begin{array}{l}\delta_k\theta^{(+)}=\Pi^+k^{(-)}, \qquad
\delta_kx^-=0,\\
\delta_kx^+={\rm i}\delta_k\theta^{(+)}\theta^{(+)}, \qquad \delta_ke =
2{\rm i}ek^{(-)}\dot\theta^{(+)}.\end{array}
\eqno{(4)}$$
An interesting property of the superparticle in (1,0) flat superspace
is off-shell closure of gauge algebra for the model (it is not so for
another dimensions). To check this assertion, consider the following
transformations with a bosonic parameter $\xi^-$
$$
\delta_\xi x^-=-\xi^-\Pi^+{\dot x}^-, \qquad \delta_\xi e = \xi^-e^2
(e^{-1}\Pi^+)^{\cdot}.
\eqno{(5)}$$
It is a trivial symmetry of the action in the sense that it vanishes
on-shell and, consequently, doesn't remove a number of degrees of
freedom of the theory. In the presence of this symmetry, however, the
full gauge algebra turns out to be closed and has the following form:
$$
\begin{array}{ll}{} &\qquad\alpha = 2{\rm i}k_2^{(-)}k_1^{(-)}\Pi^+\cr
[\delta_{k_1}\delta_{k_2}] = \delta_\alpha + \delta_\xi + \delta_{k_3},
&\qquad\xi^-=2{\rm i}k_2^{(-)}k_1^{(-)}\\
{}&\qquad k_3^{(-)}=2{\rm i}k_2^{(-)}k_1^{(-)}\dot\theta^{(+)};\cr
[\delta_k\delta_\xi ]=\delta_{\xi_1}, &\qquad
\xi_1^-=2{\rm i}\xi^-k^{(-)}\dot\theta^{(+)}\cr
[\delta_{\xi_1},\delta_{\xi_2}]=\delta_{\xi_3}; &\qquad
\xi_3^-=(\xi_1^-\dot\xi_2^--\xi_2^-\dot\xi_1^-)\Pi^+;\cr
[\delta_k,\delta_\alpha ]=\delta_{k_1}, &\qquad k_1^{(-)} = \alpha{\dot
k}^{(-)}-\dot\alpha k^{(-)};\cr
[\delta_\xi ,\delta_\alpha ]=\delta_{\xi_1}, &\qquad \xi_1^-=\alpha
\dot\xi^--2\xi^-\dot\alpha ;\cr
[\delta_{\alpha_1},\delta_{\alpha_2}]=\delta_{\alpha_3}, &\qquad
\alpha_3=\alpha_2\dot\alpha_1-\alpha_1\dot\alpha_2.
\end{array}
\eqno{(6)}$$
An application of the Dirac procedure to the model (1) leads to the
following constraints system
$$
P_e\approx 0, \qquad P_{(+)}+{\rm i}\theta^{(+)}P_+\approx 0, \qquad
P_+P_- \approx 0
\eqno{(7)}$$
where ($P_e,P_+,P_-,P_{(+)}$) are canonically conjugate momenta for the
variables ($e,x^+,x^-,\theta^{(+)}$) respectively. Specific feature of
$d=2$ lies in the fact that the last condition implies two
possibilities:\\
a) $P_+=0,~P_-\neq 0$ and, consequently, all constraints are first
class;\\
b) $P_-=0,~P_+\neq 0$ whence we conclude that $P_e\approx 0$ and $P_-
\approx 0$ are first class while $P_{(+)}+{\rm i}\theta^{(+)}P_+
\approx 0$ is second class.

\section{The model in a curved background}

Introducing a set of basis one-forms $e^A={\rm d}z^M{e_M}^A$, where
${e_M}^A$ is the supervielbein, the extension of the (1,0) superparticle
action to curved superspace can be written in the form
$$
S=\int{\rm d}\tau\,e^{-1}\,{\dot z}^M{e_M}^+{\dot z}^N{e_N}^-
\eqno{(8)}$$
where $M=(m,(+))$ are coordinate indices and $A=(+,-,(+))$ are tangent
space indices of (1,0) curved superspace. World indices appear on the
coordinates $z^M=(x^m,\theta^{(+)})$ and supervielbein only.

It should be noted that in the presence of an arbitrary background
superfield the action (8) doesn't possess any local symmetry except
the reparametrization invariance ($\delta_\alpha e =(\alpha
e)^{\cdot}$, $\delta_\alpha z^M=\alpha{\dot z}^M$). Thus it is not the
superparticle model yet. Let us find what constraints on background
geometry follow from the requirement of correct number of degrees of
freedom for the theory or, what is the same, from the requirement of
correct number of first and second class constraints for the model.

The momenta conjugate to $e$ and $z^M$ are
$$
P_e = 0, \qquad P_M=e^{-1}{e_M}^+{\dot z}^N{e_N}^- +e^{-1}{e_M}^-{\dot
z}^N{e_N}^+
\eqno{(9)}$$
whence one gets the primary constraints
$$
P_e\approx 0, \qquad P_{(+)}\approx 0
\eqno{(10)}$$
where we denoted $P_A\equiv {e_A}^MP_M$.

The canonical Hamiltonian is given by
$$
H=P_e\lambda_e+eP_+P_- + \lambda^{(+)}P_{(+)}
\eqno{(11)}$$
where $\lambda_e$, $\lambda^{(+)}$ are Lagrange multipliers enforcing
the constraints (10). The graded Poisson bracket has the form
$$
\{AB\}=(-)^{\epsilon_A\epsilon_N}\frac{\vec\partial A}{\partial z^N}~
\frac{\vec\partial B}{\partial P^N}-(-)^{\epsilon_A\epsilon_B +
\epsilon_B\epsilon_N}\frac{\vec\partial B}{\partial z^N}~
\frac{\vec\partial A}{\partial P^N}
\eqno{(12)}$$
where $\epsilon_A$ is a parity of a function $A$. It is straightforward
to check that
$$
\{P_A,P_B\}={T_{AB}}^CP_C-\left({\omega_{AB}}^C-(-)^{\epsilon_A
\epsilon_B}{\omega_{BA}}^C\right)P_C
\eqno{(13)}$$
where ${T_{AB}}^C$ are components of the torsion two-form
$$
{T_{BC}}^A=(-)^{\epsilon_B(\epsilon_M+\epsilon_C)}{e_C}^M{e_B}^N\left(
\partial_N{e_M}^A+(-)^{\epsilon_N(\epsilon_M+\epsilon_B)}{e_M}^B
{\omega_{NB}}^A -\right.
$$
$$
\qquad{}\left.-(-)^{\epsilon_N\epsilon_M}(\partial_M{e_N}^A +
(-)^{\epsilon_M(\epsilon_N+\epsilon_B)}{e_N}^B{\omega_{MB}}^A)\right)
\eqno{(14)}$$
and ${\omega_{MA}}^B$ are components of the superconnection one-form
which, in the case of $d=2$, can be chosen in the form [12]
$$
{\omega_{NA}}^B=\omega_N{\delta_A}^BN_B, \qquad N_B=(N_+,N_-,N_{(+)}) =
(1,-1,1/2).
\eqno{(15)}$$
The preservation in time of the primary constraint $P_e\approx 0$ leads
to the secondary one
$$
P_+P_-\approx 0.
\eqno{(16)}$$
As in flat case, this condition implies two possibilities which should
be considered separately.

Because inclusion of interaction has to preserve dynamical contents of
a theory (a number of degrees of freedom), in the first case we must
demand all constraints to be first class. It means that all brackets
between the constraints have to vanish weakly. Taking into account Eq.
(15) and the fact that ${T_{AB}}^C=-(-)^{\epsilon_A\epsilon_B}
{T_{BA}}^C$, it is easy to get the following weak equations
$$
\begin{array}{c}
\{P_+,P_+\}=0,\qquad\{P_+,P_{(+)}\}\approx T_{+(+)}{}^-P_-, \\
\{P_{(+)},P_{(+)}\}\approx T_{(+)(+)}{}^-P_-\end{array}
\eqno{(17)}$$
whence we conclude that it is necessary to require the following
constraints on background
$$
{T_{+(+)}}^-=0, \qquad {T_{(+)(+)}}^-=0.
\eqno{(18)}$$
In the second case, the constraints $P_e\approx 0$, $P_-\approx 0$ are
to be first class while the constraint $P_{(+)}\approx 0$ is to be
second class. Evaluating brackets between the constraints
$$
\begin{array}{c}
\{P_-,P_-\}=0,\qquad\{P_-,P_{(+)}\}\approx T_{-(+)}{}^+P_+, \\
\{P_{(+)},P_{(+)}\}\approx T_{(+)(+)}{}^+P_+\end{array}
\eqno{(19)}$$
we conclude that they have correct class if the following conditions:
$$
{T_{-(+)}}^+=0, \qquad {T_{(+)(+)}}^+=-2{\rm i}\Phi
\eqno{(20)}$$
are fulfilled, where $\Phi$ is a nonvanishing everywhere arbitrary
superfield (the multiplier -2i is taken for further convenience).

Thus, a correct formulation for the (1,0) superparticle model in a
curved background implies Eqs. (18) and (20).

Now, let us reconstruct local transformations which are consistent
with Eqs. (18), (20) in the Lagrangian formalism. The following remark
will be important here. Since ${e_M}^A(z)$ and $\omega_A(z)$ are background
superfields, they transform under arbitrary variation $\delta z^M$ only
through their coordinate dependence ($\delta{e_M}^A=\delta z^N\partial_N
{e_M}^A$). Thus, local Lorentz and gauge transformations are not
consistent. Following the standard arguments [12], we accompany $\delta
z^N$ by compensating local Lorentz transformation $\delta {e_M}^A=
\delta z^N\partial_N{e_M}^A+\delta_L{e_M}^A$ with the $\delta
z^N$-dependent parameter $L$ such that the resulting $\delta{e_M}^A$ is
Lorentz vector. The results are
$$
\begin{array}{l}\delta{e_M}^A=\delta z^N{\rm D}_N{e_M}^A\equiv \delta
z^N\left(\partial_N{e_M}^A+(-)^{\epsilon_N(\epsilon_M+\epsilon_C)}
{e_M}^C{\omega_{NC}}^A\right),\\
\delta\omega_A=\delta z^N{e_N}^B{\rm D}_B\omega_A-\partial_A (\delta
z^N{e_N}^B\omega_B).\end{array}
\eqno{(21)}$$
Armed with this note, we have for any variation $\delta z^N$
$$
\delta ({\dot z}^A)={\cal D}(\delta z^A)-\delta z^C{\dot z}^B{T_{BC}}^A
\eqno{(22)}$$
where ${\dot z}^A\equiv{\dot z}^N{e_N}^A$, $\delta z^A\equiv \delta z^N
{e_N}^A$ and ${\cal D}(\delta z^A)$ is a covariantized derivative
$$
{\cal D}(\xi^A)=\dot\xi^A+\xi^B{\dot z}^M{\omega_{MB}}^A.
\eqno{(23)}$$
Now, it is straightforward to check that the most general (modulo
$\alpha$-reparametrizations) transformations leaving Eq. (8)
invariant and being consistent with Eqs. (18) and (20) have the form
$$
\begin{array}{l}\delta_kz^M{e_M}^{(+)}={\dot z}^M{e_M}^+k^{(-)},\cr
\delta_kz^M{e_M}^a=0,\cr
\delta_ke=2{\rm i}e\Phi k^{(-)}{\dot z}^M{e_M}^{(+)} - e\delta_kz^M
{e_M}^{(+)}(T_{+(+)}{}^++T_{-(+)}{}^-).\end{array}
\eqno{(24)}$$

Thus, Eqs. (24) present the most general form of Siegel's $k$-symmetry
for the (1,0) superparticle in (1,0) curved superspace. Note that these
transformations contain nontrivial contributions including the torsion
superfield and don't coincide with the direct generalization of the
flat Siegel's transformations.

The equations of motion for the (1,0) superparticle in a curved
background have the form
$$
\begin{array}{l}{\cal D}(e^{-1}{\dot z}^-)+e^{-1}{\dot z}^-{\dot z}^B
{T_{B+}}^+ +e^{-1}{\dot z}^+{\dot z}^B{T_{B+}}^-=0,\\
{\cal D}(e^{-1}{\dot z}^+)+e^{-1}{\dot z}^-{\dot z}^B
{T_{B-}}^+ +e^{-1}{\dot z}^+{\dot z}^B{T_{B-}}^-=0,\\
{\dot z}^-{\dot z}^B{T_{B(+)}}^++{\dot z}^+{\dot z}^B{T_{B(+)}}^-=0,\\
{\dot z}^M{e_M}^+{\dot z}^N{e_N}^-=0\end{array}
\eqno{(25)}$$
and it is implied that the conditions (18) or (20) (in accordance with
the case) are fulfilled.

\section{Gauge algebra and the full set of (1,0) supergravity
constraints}

As was shown in Sec. 2, the full gauge algebra for the (1,0)
superparticle in flat superspace turned out to be closed. What happen
with the algebra when the model is considered in a curved background?
Introducing the following transformations with a bosonic parameter
$\xi^-$
$$
\begin{array}{l}
\delta_\xi z^M=-\xi^-{\dot z}^+{\dot z}^-{e_-}^M,\\
\delta_\xi e=\xi^-e^2{\cal D}(e^{-1}{\dot z}^+)+\xi^-e{\dot z}^+{\dot z}^-
{T_{+-}}^+ +\\
+\xi^-e{\dot z}^+{\dot z}^{(+)}{T_{(+)-}}^- + \xi^-e{\dot z}^+{\dot
z}^+{T_{+-}}^-,\end{array}
\eqno{(26)}$$
it is straightforward to check that the commutator of two
$k$-transforma\-tions has the form
$$
[\delta_{k_1},\delta_{k_2}]=\delta_\alpha +\delta_{k_3}+\delta_\xi ,
\eqno{(27)}$$
\begin{eqnarray*}
\alpha &=& 2{\rm i}{\dot z}^+k_2^{(-)}k_1^{(-)}\Phi ,\qquad \xi^- =
2{\rm i}k_2^{(-)}k_1^{(-)}\Phi ,\\
k_3^{(-)}&=&2{\rm i}k_2^{(-)}k_1^{(-)}{\dot z}^{(+)}\Phi - 2 k_2^{(-)}
k_1^{(-)}{\dot z}^+{T_{+(+)}}^+ -\\
{}&-& k_2^{(-)}k_1^{(-)}{\dot z}^+{T_{(+)(+)}}^{(+)}.
\end{eqnarray*}
To get Eq. (27) it is necessary to use the constraints (18), (20) and
the consequences of the Bianchi identities (which are solved in the
presence of (18), (20))
$$
\begin{array}{l}
2{\rm D}_{(+)}[{T_{+(+)}}^++{T_{-(+)}}^-]+2{\rm i}\partial_+\Phi +2{\rm
i}\Phi [{T_{+-}}^-+2{T_{+(+)}}^{(+)}]+\\
\quad{}+{T_{(+)(+)}}^{(+)}[{T_{+(+)}}^+ +
{T_{-(+)}}^-]=0,\\
\partial_-\Phi +\Phi [{T_{+-}}^++2{T_{-(+)}}^{(+)}]=0,\\
\partial_{(+)}\Phi +\Phi [{T_{+(+)}}^++{T_{(+)(+)}}^{(+)}]=0.
\end{array}
\eqno{(28)}$$
Taking into account Eqs. (25), it is easy to check that the
transformations (26) are trivial symmetry of the action (8) without
imposing of new constraints on background. The rest commutators in the
algebra are
$$
\begin{array}{l}[\delta_{\xi_1},\delta_{\xi_2}]=\delta_{\xi_3},\qquad \xi_3^-=
(\xi_1^-\dot\xi_2^--\xi_2^-\dot\xi_1^-){\dot z}^+,\cr
[\delta_k,\delta_\xi ]=\delta_{k_1}+\delta_{\xi_1},\cr
k_1^{(-)}=\xi^-{\dot z}^+{\dot z}^-k^{(-)}[{T_{(+)-}}^{(+)}-
{T_{+-}}^+],\cr
\xi_1^-=2{\rm i}\Phi\xi^-k^{(-)}{\dot z}^{(+)}-\xi^-k^{(-)}{\dot z}^+
{T_{+(+)}}^+,\end{array}
\eqno{(29)}$$
where we have used the constraints (18), (20) and the consequences of
the Bianchi identities
\begin{eqnarray*}
R_{+(+)}&=&{\rm D}_{(+)}{T_{+-}}^-+{\rm D}_+{T_{-(+)}}^-
+{T_{(+)+}}^{(+)}{T_{(+)-}}^-+{T_{(+)+}}^+{T_{+-}}^-,\\
R_{(+)-}&=&{\rm D}_{(+)}{T_{-+}}^++{\rm D}_-{T_{+(+)}}^+
+{T_{(+)-}}^{(+)}{T_{(+)+}}^+-\\
{}&-&2{\rm i}\Phi{T_{-+}}^{(+)} +{T_{(+)-}}^-{T_{-+}}^+,&(30)\cr
R_{(+)(+)}&=&2{\rm i}\Phi{T_{+-}}^--2{\rm D}_{(+)}{T_{(+)-}}^--
{T_{(+)(+)}}^{(+)}{T_{(+)-}}^-,\\
\partial_-\Phi &+& \Phi [{T_{+-}}^++2{T_{-(+)}}^{(+)}]=0.
\end{eqnarray*}
Thus, the full gauge algebra turns out to be closed and nontrivially
deformed as compared to the flat one (6).

As was noted above, the set (18), (20) is minimal one which implies
correct inclusion of interaction. In general case, we can add some
additional constraints to Eqs. (18), (20) and the resulting model will
be the superparticle in the presence of more rigid (restricted)
background. Let us examine, now, one of such possibilities. Consider
the transformations, which are direct generalization of the flat one
$$
\begin{array}{l}
\delta_kz^M{e_M}^{(+)}={\dot z}^M{e_M}^+k^{(-)},\\
\delta_kz^M{e_M}^a=0,\\
\delta_ke=2{\rm i}ek^{(-)}{\dot z}^M{e_M}^{(+)};\end{array}
\eqno{(31)}$$
$$
\begin{array}{l}\delta_\xi z^M{e_M}^-=-\xi^-{\dot z}^M{e_M}^+{\dot z}^N
{e_N}^-,\\
\delta_\xi e=\xi^-e^2{\cal D}(e^{-1}{\dot z}^M{e_M}^+).\end{array}
\eqno{(32)}$$
The requirement of the invariance of the action (8) under Eq. (31)
leads to the following constraints on a background:
$$
\begin{array}{l} {T_{(+)(+)}}^+=-2{\rm i}, \qquad {T_{(+)(+)}}^-=
{T_{+(+)}}^- = {T_{-(+)}}^+ =0,\\
{T_{+(+)}}^+ + {T_{-(+)}}^-=0\end{array}
\eqno{(33)}$$
while invariance under Eq. (32) implies the conditions
$$
{T_{(+)-}}^- = {T_{(+)-}}^+ = {T_{+-}}^+ = {T_{+-}}^- =0.
\eqno{(34)}$$
Taking into account the symmetry properties of the torsion, the
constraints (33) and (34) can be written in the following compact form
$$
{T_{(+)(+)}}^a = -2{\rm i}\delta^{a+}, \qquad {T_{ab}}^c = {T_{a(+)}}^b
=0.
\eqno{(35)}$$
Because the conditions (18), (20) are present among Eqs. (35), the
model (8), (31) and (32) is the superparticle. Using the Bianchi
identities, it is easy to show that the set (35) is equivalent
to the full set of (1,0) supergravity constraints [12]. In the
presence of Eqs. (35) the equations of motion for the model are written
in the form
$$
\begin{array}{l}{\dot z}^M{e_M}^+{\dot z}^N{e_N}^-=0,\qquad {\cal D}
(e^{-1}{\dot z}^N{e_N}^+)=0,\\
{\dot z}^N{e_N}^-{\dot z}^M{e_M}^{(+)}=0, \qquad {\cal D}
(e^{-1}{\dot z}^N{e_N}^-)=0,\end{array}
\eqno{(36)}$$
and, consequently, the $\xi^-$-symmetry is absent on-shell just as in
flat case. The full gauge algebra of the theory turns out to be closed
and not deformed as compared to the flat one
$$
\begin{array}{ll}[\delta_{k_1}\delta_{k_2}]=\delta_\alpha +
\delta_{k_3} + \delta_\xi, &\quad \alpha = 2{\rm
i}k_2^{(-)}k_1^{(-)}{\dot z}^+,\cr
{}&\quad k_3^{(-)}=2{\rm i}k_2^{(-)}k_1^{(-)}{\dot z}^{(+)},\cr
{}&\quad \xi^-=2{\rm i}k_2^{(-)}k_1^{(-)}\cr
[\delta_k\delta_\xi ]=\delta_{\xi_1}, &\quad
\xi_1^-=2{\rm i}k^{(-)}\xi^-{\dot z}^{(+)}\cr
[\delta_{\xi_1},\delta_{\xi_2}]=\delta_{\xi_3}, &\quad
\xi_3^-=(\xi_1^-\dot\xi_2^--\xi_2^-\dot\xi_1^-){\dot z}^+.\end{array}
\eqno{(37)}$$
Thus, the full set of supergravity constraints follows from the
requirement that the direct generalization of the flat algebra (31), (32)
is realized in the model. In this case, the algebra is not deformed as
compared to the flat one. One can note the similarity of this approach
with the light-like integrability conditions of Ref. [9].

\section{Conclusion}

In the present paper we have considered the (1,0) superparticle model
in (1,0) curved superspace. It was shown that correct inclusion of
interaction implies some set of constraints on the background which, in
general case, doesn't coincide with the full set of (1,0) supergravity
constraints. Lagrangian gauge transformations which are consistent with
these constraints contain nontrivial contributions including the torsion
superfield and don't coincide with the direct generalization of the
flat gauge transformations to a curved background. Gauge algebra of the
theory was shown to be closed and nontrivially deformed as compared to
the flat one. The requirements leading to the full set of (1,0)
supergravity constraints were proposed.

As was shown in Sec. 2, full gauge algebra of the theory turned out to
be closed. In the case of another dimensions the algebra is open. In
the first order formalism, however, the algebra turns out to be closed
again [13], and one can directly generalize the analysis of the
present paper to the case of another dimensions. For example, in the
case of (1,1) superspace the situation looks quite analogously and the
results will be published later. We hope as well, that similar analysis
can be fulfilled for the superstring case too.

\bigskip

\centerline{Acknowledgments}

\bigskip

This work is supported in part by ISF Grant No M2I000 and
European Community Grant No INTAS-93-2058.

\bigskip

\centerline{References}

\bigskip

\noindent
1. L. Brink and J. Schwarz, Phys. Lett. B {\bf 100}, 310 (1981).\\
2. M. Green and J. Schwarz, Phys. Lett. B {\bf 136}, 367 (1984).\\
3. E. Witten, Nucl. Phys. B {\bf 266}, 245 (1986).\\
4. J.A. Shapiro and C.C. Taylor, Phys. Lett. B {\bf 181}, 67 (1986);
{\bf 186}, 69 (1987).\\
5. M.T. Grisaru, P. Howe, L. Mezincescu, B.E.W. Nilsson and P.K.
Townsend, Phys. Lett. B {\bf 162}, 166 (1985).\\
6. J.J. Atick, A. Dhar and B. Ratra, Phys. Lett. B {\bf 169}, 54
(1986).\\
7. E. Bergshoeff, E. Sezgin and P.K. Townsend, Phys. Lett. B {\bf 169},
191 (1986).\\
8. W. Siegel, Phys. Lett. B {\bf 128}, 397 (1983).\\
9. E. Bergshoeff, P.S. Howe, C.N. Pope, E. Sezgin and E. Sokatchev,
Nucl. Phys. B {\bf 354}, 113 (1991).\\
10. A.A. Deriglazov, Int. J. Mod. Phys. A {\bf 8}, 1093 (1993).\\
11. L. Brink, M. Henneaux and C. Teitelboim, Nucl. Phys. B {\bf 293},
505 (1987).\\
12. M. Evans, J. Louis and B.A. Ovrut, Phys. Rev. D {\bf }, 3045
(1987).\\
13. G. Sierra, Class. Quant. Grav., {\bf 3}, L67 (1986).
\end{document}